\documentclass[aps,prl,twocolumn,showpacs,preprintnumbers,
               amsmath,amssymb]{revtex4}
\usepackage{graphicx}
\def\Comment#1{}
\newcommand{\bean}{\begin{eqnarray*}}
\newcommand{\eean}{\end{eqnarray*}}

\newcommand{\gapproxeq}{\lower
.7ex\hbox{$\;\stackrel{\textstyle >}{\sim}\;$}}
\newcommand{\lapproxeq}{\lower
.7ex\hbox{$\;\stackrel{\textstyle <}{\sim}\;$}}

\newcommand\lsim{\mathrel{\rlap{\lower4pt\hbox{\hskip1pt$\sim$}}
    \raise1pt\hbox{$<$}}}
\newcommand\gsim{\mathrel{\rlap{\lower4pt\hbox{\hskip1pt$\sim$}}
    \raise1pt\hbox{$>$}}}
\newcommand{\ba}{\begin{array}}
\newcommand{\ea}{\end{array}}
\newcommand{\nn}{\nonumber}

\newcommand{\be}{\begin{equation}}
\newcommand{\ee}{\end{equation}}
\newcommand{\bear}{\begin{eqnarray}}
\newcommand{\eear}{\end{eqnarray}}

\newcommand{\ket}{\,\rangle}
\newcommand{\bra}{\langle \,}

\newcommand{\cO}{{\cal O}}

\newcommand{\mA}{\mathcal{A}}

\newcommand{\mF}{\mathcal{F}}

\newcommand{\mM}{\mathcal{M}}
\newcommand{\mO}{\mathcal{O}}
\newcommand{\mR}{\mathcal{R}}

\newcommand{\Frac}[2]{\frac{\displaystyle #1}{\displaystyle #2}}
\newcommand{\Int}{\displaystyle{\int}}

\begin{document}


\title{
On the structure of two-point  Green-functions\\
at next-to-leading order in $1/N_C$
}

\author{Juan Jos\'e Sanz-Cillero
}

\affiliation{
Department of Physics, Peking
University, Beijing 100871, P.R. China}

\email{cillero@th.phy.pku.edu.cn}

\date{\today}

\begin{abstract}
The structure of the two-point QCD Green-functions is studied in this note
in the limit of large number of colours.
Their general form at next-to-leading order in $1/N_C$ is derived
keeping the infinite resonance summation and without relying on
a particular realization of the hadronic action.
It is found that the contributions from chiral operators
without resonance fields  of order $p^4$ or higher
are irrelevant for the computation of the correlators
and, hence, they can be dropped at the beginning of the calculation.
The possibility of a more general cancelation of these local terms
at the level of the generating functional is discussed.
\end{abstract}\vskip .5cm


\maketitle


\section*{A large--$N_C$ hadronic field theory}

Quantum Chromodynamics (QCD) is  greatly  simplified
in 't~Hooft's large--$N_C$ limit~\cite{NC1,NC2,NC3}. Quark loops and non-planar topologies
become suppressed and the theory can be then expressed in terms of tree-level hadron
exchanges.
In order to produce the perturbative QCD logarithms at deep euclidian momenta,
an infinite number of narrow-width states is needed.

Meson loops are absent at large--$N_C$ and  the amplitudes are
analytic but for the real poles related to the narrow-width
intermediate states.
It is then possible to
write down a lagrangian containing an infinite set of hadronic
fields $\Phi_k$ of mass $M_k$ that do not interact each other when
$N_C\to \infty$. In this note, we focus our attention on the meson
sector, containing the Goldstone modes from the spontaneous chiral
symmetry breaking and the mesonic resonances. The main assumption in
the present work is that the vertex functions $\Gamma^{(n)}_{\Phi_1
...\Phi_n}$  can be described at the leading order in $1/N_C$ (LO)
by a finite number of hadronic operators.  These must follow the expected   scaling
$N_C^{1-\frac{n}{2}}$~\cite{NC3}, although the $N_C$ dependence
may not appear explicitly but hidden in the couplings $\lambda^{\Phi_1...\Phi_n}$
of the corresponding operators.

As result of such resonance lagrangian, loops arise at subleading  orders.
The $N_C$--scaling of the operators in the LO action $S_0^{\rm LO}[\Phi_k,J]$
converts its loop expansion~\cite{Ryder}
into an expansion in powers of $1/N_C$.
The one-loop amplitudes  contain ultraviolet (UV) divergences
and next-to-leading order (NLO) counter-terms can be required
to fulfill the renormalization.

This work studies the two-point correlators
$\bra T\{ \bar{q}\Gamma q(x)\, \, \bar{q}\Gamma q(0)\}\ket$
or combinations of them, where $\bar{q}\Gamma q$ refers to some
vector or axial-vector current, or to scalar or pseudo-scalar densities.
These Green-functions are provided by the corresponding  scalar functions $\Pi(s)$
and their short-distance behaviour  can be derived through
the operator product expansion (OPE)~\cite{OPE}.
The aim of this article is to show the structure of UV divergences and counter-terms of $\Pi(s)$
at NLO in $1/N_C$.
In order to avoid any model dependence, the infinite tower of states is kept all along the
paper. Likewise, no explicit realization of the lagrangian is assumed,
just the existence of such hadronic action.

Phenomenological large-$N_C$ analyses
have found that some $\cO(p^4)$ chiral operators without resonance fields
are absent from the hadronic lagrangian both at LO and at NLO in $1/N_C$~\cite{vanishing}.
We will often  refer along the paper to the chiral order $\cO(p^m)$  of the operators
in the way of the chiral counting, with $p^m$ denoting the
number or derivatives or external auxiliary fields $v^\mu, a^\mu, s, p$~\cite{WE:79,chpt}.
This note  proves that even if
the presence of these chiral operators is needed to renormalize
some vertex functions, they turn out to be irrelevant for
the calculation of $\Pi(s)$
and do not have physical content~\cite{EoM-chpt}.
The correlators are shown to be fully described in terms of
the couplings in  the LO action.
A change in $\alpha_s$ would modify the hadronic parameters but $\Pi(s)$ would be
still determined by the operators in  $S^{\rm LO}_0[\Phi_k,J]$;
new operators could arise at NLO but the correlator
can be completely recovered even if we ignore the value of these new couplings.

The functions $\Pi(s)$ are analytical in the variable $s$ out of
the positive real $s$--axis and they accept general complex variable
techniques~\cite{analyticity,sumrules3,sumrules4,moments}.
Based on the asymptotic behaviour at
$|s|\to\infty$, prescribed by the OPE, one can write
$m$--subtracted dispersion relations for $m\geq n$, where $n$ denotes the minimal number
of subtractions required for $\Pi(s)$.
Other quantities with these properties
are the two-meson form factors that  vanish at short distances and accept unsubtracted dispersion
relations~\cite{Brodsky,SSPPrcht,tesis-Natxo}. Nevertheless, they bring further complications
on other aspects and they will not be discussed here.

\section*{Moments and $n$--subtracted relations}

For the Green-function analysis, we define a set of  moments $\mA_\Pi^{(m)}(s)$,
which can be related to the spectral function for $m\geq n$~\cite{moments}:
\begin{equation}
\mA_\Pi^{(m)}(s)\equiv \Frac{(-s)^m}{m!}\,\Frac{d^m}{ds^m}\Pi(s)
=\displaystyle{\int} \, \Frac{(-s)^m \,\,dt}{(t\, -\,
s)^{m+1}} \,\,\Frac{1}{\pi}\mbox{Im}\Pi(t)\, .
\end{equation}
The moments with $m<n$ can be derived through the recursive relation
\begin{equation}
\label{eq.matching}
\Frac{\mA_\Pi^{(m-1)}(s)}{s^{m-1}}\quad =\quad \Frac{\mA_\Pi^{(m-1)}(s_{0})}{s_{0}^{m-1}}
\, - \, m \, \Int^{s}_{\hspace*{-0.15cm} s_{0}} \, \,\Frac{dt}{t^m}\, \mA_\Pi^{(m)}(t)\, .
\end{equation}
In order to fix $\Pi(s)=\mA_\Pi^{(0)}(s)$ one needs to provide the value of $n$ subtraction
constants, $\{\mA_\Pi^{(m)}(s_0)\}_{m=0}^{n-1}$ at some energy $s_0$.

In some cases, the OPE may prescribe $\Pi(s)\stackrel{s\to -\infty}{\longrightarrow}0$ but
the unsubtracted dispersive relation leads to an infinite resonance summation
${\Pi(s)=\sum_{\Phi_k} \Pi_{\Phi_k}(s)}$
that is badly defined~\cite{L8Peris,PIalpha}.
For instance, the unsubtracted dispersion relation for the  SS+PP
correlator is not convergent. Hence, the infinite resonance summation in
$\Pi(s)_{_{\rm SS-PP}}$ is not absolutely convergent~\cite{L8Peris}.
One needs to perform a number of subtractions $n$
higher than what is expected just from OPE arguments.
%
%
Once the infinite series has been summed up in $\mA_\Pi^{(n)}(s)$,
one can use Eq.~(\ref{eq.matching}) to recover
the lower moments $\{\mA_\Pi^{(m)}(s)\}_{m=0}^{n-1}$.
The choice $s_0=-\infty$ gives $\mA_\Pi^{(m)}(s_0)=0$ for
the two-point Green-functions that are chiral order-parameters~\cite{OPE} and fixes $\Pi(s)$.

The Green-functions that are not chiral order-parameters are more cumbersome.
The subtraction constants can be obtained from
the renormalized perturbative QCD correlator $\Pi(s_0)^{^{\rm pQCD}}$ and
its moments $\mA_\Pi^{(m)}(s_0)^{^{\rm pQCD}}$
at deep euclidian momentum ${   -s_0\gg \Lambda_{\rm QCD}^2   }$.
Furthermore, $\mA_\Pi^{(m)}(s_0)$ is a finite and  pure QCD quantity  for $m\geq n$
that can be fully determined in the resonance theory.
Its matching to perturbative QCD
allows to recover $\alpha_s$ in terms of the hadronic parameters~\cite{PIalpha}.
In the chiral limit,
provided the value of the strong coupling constant and one renormalization condition
for $\Pi(s_0)^{^{\rm pQCD}}$ (not fixed by QCD),
the rest of moments $\mA_\Pi^{(m)}(s_0)^{^{\rm pQCD}}$
turns out to be
completely  determined.
This demonstrates that all the QCD information of $\Pi(s)$ is actually contained in
$\mA_\Pi^{(n)}(s)$.
Moreover, this moment, determined within the resonance $1/N_C$ framework,
yields the evolution from $s_0$ in
the perturbative QCD regime down to any energy.

Before entering into the  calculation of $\Pi(s)$, some more preliminary definitions
are  needed.
Let us consider a general function $f(t)$, analytical in the whole complex plane except for
logarithmic branches and single and double poles at $t=M_k^{r\,\, 2}$.
If $f(t)$ accepts a $n$--subtracted dispersive relation, one has the mathematical identity
for its moment,
%
%
\begin{eqnarray}  \label{eq.master}
\mA_f^{(n)}(s)
\, &=& \, \Delta \mA_f^{(n)}(s)
\\ &&  \hspace*{-0.5cm}+
\sum_k\left[ \Frac{ (-s)^n\, Z_k^f}{(M_k^{r\,\,2}-t)^{n+1}}
    + \Frac{(n+1) (-s)^n \,  D_k^f}{(M_k^{r\,\,2}-t)^{n+2}}  \right] , \nn
\end{eqnarray}
with the coefficients
\begin{eqnarray}
D_k^f  &=&  \lim _{t\to M_k^{r\, \,2} }\, \mbox{Re}\left\{  (t-M_k^{r\,\,2})^2\, f(t)  \right\}\, ,
\label{eq.D} \nn \\
Z_k^f &=& \, -\, \lim_{t\to M_k^{r\,\,2} }\, \mbox{Re}\left\{
                \Frac{d}{dt}\left[  (t-M_k^{r\,\,2})^2\, f(t)\right]   \right\}\, .
\label{eq.C}
\end{eqnarray}
and the absorptive contribution
\begin{equation}
\begin{array}{l}
\Delta\mA_f^{(n)}(s)\, =\, \displaystyle{ \lim_{\epsilon\to 0^+}  }
\left\{
    \Frac{1}{\pi}\, \displaystyle{ \Int_{\mR_\epsilon}}
    \, \Frac{dt\,\,  (-s)^n \, \, \mbox{Im}f(t) }{(t\,-\, s)^{n+1}}\,
\right.
\\
\left. \qquad \qquad
    - \, \Frac{2}{\pi \epsilon}\,
        \displaystyle{\sum_{k}\lim_{t\to M_k^{r\,\,2}}}
        \Frac{(-s)^n\, \, (t-M_k^{r\,\,2})^2 \,\,  \mbox{Im}f(t)}{(t-s)^{n+1}}
\right\}\, ,
\end{array}\label{eq.Deltafn}
\end{equation}
with $\mR_\epsilon=[0,+\infty)-\bigcup_k (M_k^{r\,\, 2}-\epsilon,M_k^{r\,\,2}+\epsilon)$.
The function $\Delta \mA_f^{(n)}(s)$ only depends on Im$f(t)$ at $t\neq M_k^{r\,\,2}$.

For sake of clarity, the explicit derivation is shown for correlators obeying unsubtracted
dispersion relations. In order to get the $n$--subtracted expressions, one should make the replacements,
\begin{eqnarray}
\Pi(s)&\longrightarrow& \mA_\Pi^{(n)}(s)\, , \nn
\\
\Frac{1}{M_k^{r\,\,2}-s} &\longrightarrow& \Frac{(-s)^n}{(M_k^{r\,\,2}-s)^{n+1}} \, , \nn
\\
\Frac{1}{(M_k^{r\,\,2}-s)^2} &\longrightarrow& \Frac{(n+1)\, (-s)^n}{(M_k^{r\,\,2}-s)^{n+2}} \, ,
\label{eq.transcribe}
\\
\displaystyle{ \sum_{j\geq 0}} \, c_j\, s^j  &\longrightarrow&
\displaystyle{ \sum_{j\geq n}} \,\Frac{(-1)^n\, j!}{n!(j-n)!}\,  c_j\, s^j\, .  \nn
\end{eqnarray}

\section*{Large--$N_C$ amplitude}

In the large $N_C$ limit, the absorptive part of $\Pi(t)$ is composed by a series delta
functions,
\be
\Frac{1}{\pi}\mbox{Im}\Pi(t)
\,\, = \,\,
\sum_k  Z_k \, \,\delta(t-M_k^2) \, ,
\ee
with $Z_k$ and $M_k$ being, respectively,
the residues and the pole positions of the
QCD amplitude at large--$N_C$.
Using dispersive relations, one gets the form of the LO amplitude,
\be \label{eq.LO}
\Pi(s)  \,\, = \,\,
\sum_k \Frac{Z_k}{M_k^2\, -\, s} \, .
\ee

This amplitude can be described through the tree-level diagrams of a  lagrangian with
local operators containing the meson fields $\Phi_k$ of mass $M_k$~\cite{lagrangian}.
Depending on the realization of the hadronic lagrangian,
the leading action may contain $\cO(p^2)$ resonance operators
that give the resonance exchange~\cite{rcht},
\begin{equation}
\label{eq.res}
\Pi(s)_{\rm res}  \quad = \quad
\sum_k \Frac{F_k^2}{M_k^2\, -\, s} \, .
\end{equation}
Notice that if $\Pi(s)$ refers to the difference of two correlators,
the summation may contain $(-1)$ factors.

There may also be some local contributions of the form
\begin{equation}
\label{eq.contact}
\Pi(s)_{\rm loc} \, =\,\displaystyle{\sum_{j}}
\, c_j\, s^j\, .
\end{equation}
These operators are provided by  chiral perturbation theory
if we work within a chiral invariant framework~\cite{chpt}.
Some terms only depend on the external fields, e.g., $v_\mu$ and
$a_\mu$ in the case of the $\cO(p^4)$ operator $H_1\mO_{H_1}$~\cite{chpt}.
Nevertheless,  there can be chiral invariant operators, like $L_{10}\mO_{L_{10}}$
at $\cO(p^4)$,
that contain Goldstone fields in addition to the
contact terms with only external sources~\cite{chpt}.
They are non-trivial and contribute
to other processes $\mM_\Pi$ related to $\Pi(s)$ through chiral symmetry.

One may also think of further resonance operators  contributing at tree-level.
At LO, the only bilinear operators are the canonical kinetic term, e.g.,
${- \Phi_k^\dagger (\partial^2+M_k^2)\Phi_k}$ for spin--0 particles.
On the other hand, there can be resonance operators contributing
to the ${\bar{q}\,\Gamma q\to \Phi_k}$ vertices with a number of derivatives
higher than $\cO(p^2)$.
A term like this   would give the amplitude
\begin{equation}
\label{eq.respr}
\Pi(s)_{\rm res'}\, =\, \Frac{\lambda \,s^{j+1} }{M_k^{2}-s} \, .
\end{equation}
This can be always rewritten in the form
\begin{equation}
\label{eq.respr2}
\Pi(s)_{\rm res'}\, =\,
\Frac{\lambda \, (M_k^2)^{j+1} }{M_k^2-s}
\, +\, \displaystyle{ \sum_{j'}}\, c_{j'}^\lambda \, \, s^{j'}\, .
\end{equation}
Hence, these operators reproduce the same structures found before in Eqs.~(\ref{eq.res})
and (\ref{eq.contact}) and  do not contain further independent information.
The equations of motion (EoM) provide a deeper understanding of this simplification.
Every  spin--0 resonance obeys a classical  EoM of the form
${\partial^2 \Phi_k=-M_k^2\Phi_k + \chi[J,\pi] + \cO(\Phi^2)}$, where $\chi[J,\pi]$ stands for
operators containing only external sources and Goldstone
fields~\cite{functional,tesis-Natxo}.
The operators linear  on $\Phi_k$
and contributing to $\bar{q}\Gamma q\to \Phi_k$
can be fully transformed   thanks to the EoM into operators without resonance fields  and
terms that do not contribute to $\Pi(s)$.
The case of particles with spin different from zero is more complicate due to the
appearance of Lorentz structures in the EoM.
Nevertheless, the study of vector fields in the antisymmetric tensor
formalism casts a similar result~\cite{tesis-Natxo,VFFrcht}.
In any case, all that matters for this paper  is that
any tree-level contribution to $\Pi(s)$ at LO in $1/N_C$
can be written in the form of Eqs.~(\ref{eq.res}) and (\ref{eq.contact}), where
$F_k$, $M_k$, $c_j$ carry all the independent information.
These EoM simplifications will be also crucial in the NLO analysis.

Finally, matching the QFT and the dispersive expression in Eq.~(\ref{eq.LO}), one gets
\begin{equation}
F_k^2=Z_k\, , \qquad \qquad c_j=0\, .
\end{equation}
This can be better understood through the example of the vector correlator.  If
the spin--1 field are realized in the Proca formalism, there are no $\cO(p^2)$
resonance operators~\cite{spin1fields}. The lowest order is $\cO(p^3)$
and the correlator results
\begin{equation}
\Pi_{_{\rm VV}}(s)\, =\, \displaystyle{\sum_k}  \, \Frac{f_{V_k}^2 \, s}{M_{V_k}^2-s}
\, +\, c_0\,+ c_1\, s\, +\, ...
\end{equation}
It can be then recasted  in the form of Eqs.~(\ref{eq.res}) and (\ref{eq.contact}),
\begin{equation}
\Pi_{_{\rm VV}}(s)\, =\, \displaystyle{\sum_k}  \, \Frac{f_{V_k}^2 \, M_{V_k}^2 }{M_{V_k}^2-s}
\, +\, c'_0\,+ c'_1\, s\, +\, ...
\end{equation}
with $c'_0=c_0-\sum_k f_{V_k}^2$ and $c'_j=c_j$ in the rest of cases.
The dispersion relation sets $c'_j=0$ and the correlator becomes determined just in terms of
$M_{V_k}$ and ${  Z_k=f_{V_k}^2 M_{V_k}^2   }$, that contain all the physical information of $\Pi(s)$:
\begin{equation}
\label{eq.exampleLO}
\Pi_{_{\rm VV}}(s)\, =\, \displaystyle{\sum_k}  \, \Frac{f_{V_k}^2 \, M_{V_k}^2}{M_{V_k}^2-s} \, .
\end{equation}
The infinite summation in the $c'_j$ does not present a problem. To regularize it,
one can always construct a series of theories with a finite number of states containing the same
couplings $f_{V_k}$, $M_{V_k}$ but with $c_j$ chosen such that $c'_j=0$.
In the limit when one includes the infinite tower of resonance, such theories
recover the structure of the correlator in Eq.~(\ref{eq.LO})
prescribed by the dispersion relations.

\section*{Amplitude at next-to-leading order in $1/N_C$}

At NLO,  a shift in the lagrangian parameters is required
in order to absorb the one-loop UV divergences.
The renormalization of the $\cO(p^2)$ resonance couplings,
${M_k^2=M_k^{r\, \, 2}+\delta M_k^2}$, ${F_k^2=F_k^{r\,\,2}+\delta F_k^2}$
gives,
\begin{equation}
\label{eq.resNLO}
\Pi(s)_{\rm res}
\,= \,  \sum_k \left[ \Frac{F_k^{r\,\, 2}+\delta F_k^2}{M_k^{r\,\, 2}-s}
\, -\,   \Frac{F_k^{r\,\, 2}\, \delta M_k^2}{(M_k^{r\,\, 2}-s)^{2}} \right] \, .
\end{equation}

Although the analysis of some phenomenological lagrangians
have found that the corresponding local operators $c_j\mO_{c_j}$
vanish from the action at NLO~\cite{vanishing},
we will consider the more general case when $c_j\neq 0$ and a non-zero
contribution from $\Pi(s)_{\rm loc}$, given in Eq.~(\ref{eq.contact}), arises at NLO in $1/N_C$.

Higher $\cO(p^m)$ operators linear in $\Phi_k$ can be removed at NLO
in the same way as it was done at LO in Eqs.~(\ref{eq.respr}) and (\ref{eq.respr2}).
In addition, one may have at NLO in $1/N_C$
tree-level contributions
to the resonance self-energies. A generic operator of this kind  would yield the amplitude
\begin{equation}
\Pi(s)_{\rm res'}\, =\, \Frac{\lambda \,s^{j+2}}{(M_k^{r\,\, 2}-s)^2} \, ,
\end{equation}
where $M_k^2$ and $M_k^{r\,\,2}$
can be used indistinctly at NLO.  This can be reexpressed as
\begin{equation}
\Pi(s)_{\rm res'} =
\Frac{\lambda\, (M_k^{r\,\,2})^{j+2} }{(M_k^{r\,\,2}-s)^2}
\, - \,(j+2) \, \Frac{\lambda \,(M_k^{r\,\,2})^{j+1}}{M_k^{r\,\,2}-s}
\, + \displaystyle{ \sum_{j'}} c_{j'}^\lambda \, \, s^{j'} ,
\end{equation}
which reproduces the momentum structures in $\Pi(s)_{\rm res}$ and $\Pi(s)_{\rm loc}$.
Therefore, it does not provide any extra independent information to the correlator.
Again, the study of the leading order action EoM  gives  a deeper
understanding to this kind of simplifications~\cite{functional,tesis-Natxo,VFFrcht}.
Hence, any tree-level contribution to $\Pi(s)$ will be expressed
in the form of $\Pi(s)_{\rm loc}$ and $\Pi(s)_{\rm res}$ in
Eqs.~(\ref{eq.contact}) and~(\ref{eq.resNLO}).

\begin{figure}[t]\centering
\includegraphics[width=8cm,angle=0,clip]{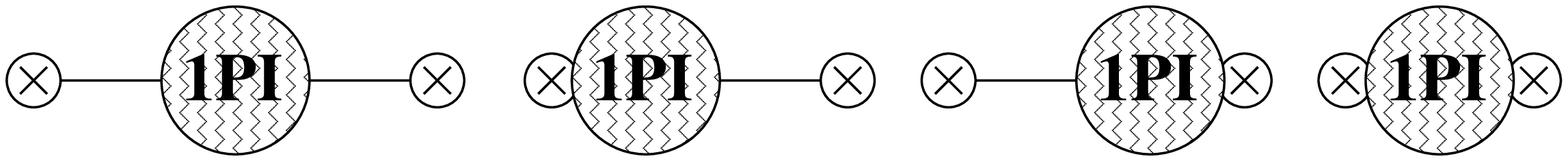}
\caption{\small 1PI contributions to
the one-loop correlators.} \label{fig.1PI}
\end{figure}

The two-point Green-functions are
provided at the one-loop level by the one-particle irreducible topologies (1PI)
shown in Fig.(\ref{fig.1PI}). Hence, the perturbative expression of $\Pi(t)$
contains single an double poles at ${  t=M_k^{r\,\, 2}  }$.
The one-loop amplitude is given by the
one- and two-point Feynman integrals, respectively $A_0(M^2)$
and  $B_0(s,M^2,M'^2)$~\cite{Passarino}.  Any term proportional to $A_0(M^2)$
is a rational functions of the form
\begin{equation}
\Pi(s)_{A_0}\, =\, \displaystyle{\sum_{j}} \, \Gamma_j^{A_0} \, s^j
+ \displaystyle{\sum_k}
\left[ \Frac{Z_k^{A_0}}{M_k^{r\,\,2}-s}
+ \Frac{D_k^{A_0}}{(M_k^{r\,\,2}-s)^{2}}\right] .
\end{equation}
The contribution from a given $\Phi_1\Phi_2$ absorptive cut is provided by the
$B_0(s,M_{\Phi_1}^2,M_{\Phi_2}^2)$ Feynman integral through the term
\begin{equation}
\label{eq.B0}
\Pi(s)_{_{\Phi_1\Phi_2}} \,=\, \xi_{\Phi_1\Phi_2}(s)\,\cdot \,  \mF_{\Phi_1\Phi_2}(s)^2
\,\cdot\, B_0(s,M_{\Phi_1}^2,M_{\Phi_2}^2)\, ,
\end{equation}
where $\mF_{\Phi_1\Phi_2}(s)$ is the $\Phi_1\Phi_2$ form-factor at large--$N_C$ and
the kinematical factor $\xi_{\Phi_1\Phi_2}(s)$ is
a known polynomial that depends on the channel under consideration~\cite{SSPPrcht,tesis-Natxo}.
For instance,
one has $\xi_{\pi\pi}(s)=\frac{2}{3}$ for the vector correlator.  The form-factor,
and more exactly the combination $\xi_{\Phi_1\Phi_2}(t)\, \mF_{\Phi_1\Phi_2}(t)^2$,  provides the
$\Phi_1\Phi_2$ contribution to the spectral function. Hence, if the correlator follows
a $n$--subtracted dispersion relation, also does the expression in Eq.~(\ref{eq.B0}).
It can be then expressed through the master relation in Eq.~(\ref{eq.master}),
\begin{equation}
\Pi(s)_{_{\Phi_1\Phi_2}}
= \Delta \Pi(s)_{_{\Phi_1\Phi_2}}
+ \displaystyle{\sum_k}
\left[ \Frac{Z_k^{^{\Phi_1\Phi_2}}}{M_k^{r\,\,2}-s}
+ \Frac{D_k^{^{\Phi_1\Phi_2}}}{(M_k^{r\,\,2}-s)^{2}}\right]  ,
\end{equation}
where $\Delta\Pi(s)_{_{\Phi_1\Phi_2}}$ is determined by Im$\Pi(s)_{_{\Phi_1\Phi_2}}$
at ${  t\neq M_k^{r\,\, 2}  }$ and it is therefore
free of UV--divergences. These are contained in the coefficients
$Z_k^{^{\Phi_1\Phi_2}}$, $D_k^{^{\Phi_1\Phi_2}}$,
defined as in Eq.~(\ref{eq.C}).  Notice that for sake of simplicity
we have written down the expression for $\Pi(s)$. The transcription
to higher moments $\mA_\Pi^{(n)}$ through Eq.~(\ref{eq.transcribe}) is left to the reader.
Putting the different one-loop diagrams together one gets the structure,
\begin{eqnarray}
\label{eq.1l}
\Pi(s)_{1\ell} &=& \Delta \Pi(s)_{1\ell}
\, + \, \displaystyle{\sum_{j}} \, \Gamma_j^{1\ell} \, s^j
\\
&&  \hspace*{-0.5cm}
+ \displaystyle{\sum_k}
\left[ \Frac{Z_k^{1\ell}}{M_k^{r\,\,2}-s}
+ \Frac{D_k^{1\ell}}{(M_k^{r\,\,2}-s)^{2}}\right]
.    \nn
\end{eqnarray}
Since the only two-meson absorptive contribution to the correlator comes from
the one-loop amplitudes,
%
%
one can write  $\Delta \Pi(s)$ instead of $\Delta \Pi(s)_{1\ell}$.
The summation of the $\Gamma_j$, $Z_k$,
$D_k$ from the different loop diagrams in $\Pi(s)_{A_0}$ and $\Pi(s)_{_{\Phi_1\Phi_2}}$
yields, respectively,
the constants $\Gamma_j^{1\ell}$, $Z_k^{1\ell}$, $D_k^{1\ell}$.
Whereas the derivation of the different terms in Eq.~(\ref{eq.1l}) is straight-forward
in the case of a QFT with a finite number of states,
its definition is more cumbersome in the infinite resonance limit.  This
discussion is relegated for the end of the section, after the main result is obtained.

%
%
%

Summing up the one-loop amplitude $\Pi(s)_{1\ell}$, the local terms $\Pi(s)_{\rm loc}$
and the resonance exchanges $\Pi(s)_{\rm res}$, one gets the correlator up to NLO:
\begin{eqnarray}
\Pi(s)&=&
\Delta\Pi(s) \, +\, \displaystyle{\sum_{j}}
\,  \left[c_j+\Gamma_j^{1\ell}\right]\,s^j
\nn
\\
&& \hspace*{-1cm}  + \displaystyle{\sum_k}
        \left[ \Frac{  Z_k}{M_k^{r\,\, 2}-s}
        + \Frac{D_k}{(M_k^{r\,\, 2}-s)^{2}}
        \right],
\label{eq.sumPI}
\end{eqnarray}
with
\begin{eqnarray}
Z_k\, &=&\,F_k^{r\,\, 2} +\delta \, F_k^2 \, + \, Z_k^{1\ell}\, ,    \nn
\\
D_k \, &=&\,- \, F_k^{r\,\,2}\, \delta M_k^2 \, +\, D_k^{1\ell}\, .
\end{eqnarray}
The coefficients $Z_k$ and $D_k$ contain UV divergences from $\Pi(s)_{1\ell}$
that are canceled through the renormalizations
\begin{eqnarray}
\delta F_k^2 &=&  \, - \, Z_k^{1\ell}\,   +\,   \Delta Z_k\, ,   \nn
\\
F_k^{r\,\, 2}\, \delta M_k^2 &=&  D_k^{1\ell}\, - \, \Delta D_k\, ,
\end{eqnarray}
where the renormalization scheme for the couplings and masses is set
by  the finite constants $\Delta Z_k$ and  $\Delta D_k$, respectively.  For instance,
one can choose the on-shell scheme $\Delta Z_k=\Delta D_k=0$.

Since the correlator $\Pi(s)$ obeys a $n$--subtracted dispersion
relation (here shown unsubtracted for sake of clarity), one can apply the identity
in Eq.(\ref{eq.master}):
\begin{equation}
\label{eq.dispPI}
\Pi(s) =\Delta \Pi(s)
+ \sum_k\left[
    \Frac{Z_k}{M_k^{r\,\,2}-s}
    +\Frac{D_k}{(M_k^{r\, \, 2}-s)^{2}}\right] .
\end{equation}
The comparison to the QFT calculation in   Eq.~(\ref{eq.sumPI})
leads to the identities
\begin{eqnarray}
F_k^{r\,\, 2}\, +\, \Delta Z_k &=& Z_k \,, \nn
\\
\Delta D_k &=& D_k \, ,
\label{eq.Lrenor}
\\
c_j \, +\, \Gamma_j^{1\ell}  &=& 0 \, . \nn
\end{eqnarray}
Provided a renormalization scheme $\Delta Z_k$, $\Delta D_k$,
the renormalized couplings $F_k^{r}$ are given by the residues of
the QCD amplitude at NLO in $1/N_C$.
The polynomial contribution vanishes identically and it is zero
for any energy, being $c_j$ and $\Gamma_j^{1\ell}$ irrelevant for the calculation of
$\Pi(s)$ (or the moment $\mA_\Pi^{(n)}(s)$ in the case of $n$-subtracted dispersion relations).
The analysis of the $SS-PP$ correlator
in the single resonance approximation provides a practical example
on the use of these techniques~\cite{SSPPrcht,tesis-Natxo}.

However, the irrelevance of  these local  terms
in other amplitudes $\mM_\Pi$ related to $\Pi(s)$ by chiral symmetry
is not so trivial.
In the case when $\Gamma_j^{1\ell}=0$, the parameters
$c_j$ become zero and the local operators $\mO_{c_j}$
disappear from the hadronic action.
This situation has been found for some matrix elements
within an approximation of large--$N_C$ QCD,
where only spin--0 states were included and
$S_0^{\rm LO}[\Phi_k,J]$ contained operators
with at most two derivatives~\cite{vanishing}.
On the other hand,
a non-zero value of $\Gamma_j^{1\ell}$ induces a coupling $c_j\neq 0$ and
the operators $c_j \mO_{c_j}$ could, at first, add some analytical
contribution to other chirally related amplitudes $\mM_\Pi$.

A last important detail is whether the summation of infinite diagrams is well defined
(as it happens for a theory with a finite number of states).
In order to regularize this limit, it is possible to follow a procedure
similar to that one employed  at LO.
One may consider the correlator $\Pi(s)^{\rm part.}$ provided by the partial summation
of a finite number of one-loop diagrams. For every partial sum one can always  keep
$F_k^{r}$ and $M_k^r$ fixed and reproducing the  residues $Z_k$ and pole
positions of the full QCD amplitude,
and with the $c_j$ couplings chosen such that  $c_j=-\Gamma_j^{1\ell,part.}$.
In the limit when one sums an infinite number of diagrams
such theories lead to the Green-function prescribed by dispersion theory in  Eq.~(\ref{eq.dispPI}).

\section*{Structure of the generating functional}

To end with, a digression on the importance of the local terms in other
amplitudes is presented. We examine the generating functional $W[J]$, given in terms of
chiral invariant operators that preserve the QCD symmetries~\cite{chpt,functional,tesis-Natxo}.
On what follows we will refer only to spin--0 correlators. Hadrons with  higher spins
complicate the argumentation since
one may find that there are no $\cO(p^2)$ resonance operators, as it happens
with the spin--1 fields in the Proca formalism~\cite{spin1fields}.

At leading order, $W[J]$ is provided  by the LO action $S_0^{\rm LO}[\Phi^{cl}_k,J]$,
with $\Phi_k^{cl}$ the classical meson fields that depend implicitly on the external
sources $J$.
Actually, the LO analysis showed that the Green-functions are determined by the
operators of lowest chiral order: $\cO(p^2)$ for the scalar correlator~\cite{rcht},
$\cO(p^3)$ for the vector correlator in the Proca formalism~\cite{spin1fields}...

Following the derivation of the  NLO amplitude in the note,
the one-loop generating functional shows the structure
\begin{eqnarray}
W[J] &=&  S_0^{\rm LO}[\Phi^{cl}_k,J]  +S^{\rm LO}_1[\Phi_k^{cl},J]
\nn \\
&& +\,  \displaystyle{\sum_{j}} c_j \,  \Int dx^d\, \mO_{c_j}[\pi^{cl},J]\,\,\,  +\,\, [...]\, ,
\end{eqnarray}
where $[...]$ stands for an irrelevant constant and operators that do not enter in
the calculation of  $\Pi(s)$
or chirally related amplitudes $\mM_\Pi$.
The term $S_1^{\rm LO}[\Phi_k^{cl},J]$ is the one-loop contribution to the generating functional
coming from the integration of the quadratic fluctuations of $\Phi_k$  in the
action $S_0^{\rm LO}[\Phi_k,J]$ around $\Phi_k^{cl}$~\cite{functional,tesis-Natxo}.
The local terms $\mO_{c_j}$ do not contain resonance fields and
only depend on external sources and Goldstone fields $\pi^{cl}$.

Although we have implicitly considered $n$--subtracted dispersion relations all along the paper, let us
suppose that the $\Pi(s)$ accept unsubtracted dispersion relations.
It is then possible to write the  decomposition,
\begin{eqnarray}
S_1^{\rm LO}[\Phi_k^{cl},J] \, &=& \,  \overline{S}_1^{\rm LO}[\Phi_k^{cl},J] \,
\\
&&\hspace*{-0.6cm}
+\,  \displaystyle{\sum_{j}} \Gamma_j^{1\ell} \, \Int \, dx^d\, \mO_{c_j}[\pi^{cl},J]
\,\,\, +\,\, [...]\, ,   \nn
\end{eqnarray}
where $\overline{S}_1^{\rm LO}[\Phi_k^{cl},J]$ generates the absorptive contribution
$\Delta \Pi(s)$
and the UV divergent terms $Z_k^{1\ell}$, $D_k^{1\ell}$. These are  renormalized
by the $\cO(p^2)$ resonance operators in $S^{\rm LO}_0[\Phi_k^{cl},J]$.
The polynomial divergences of $\Pi(s)$ are contained in the constants $\Gamma_j^{1\ell}$.
Putting the contributions    $S_0^{\rm LO}[\Phi_k^{cl},J]$
and $S_1^{\rm LO}[\Phi_k^{cl},J]$   together with the $\Pi(s)$ constraint
$c_j=-\Gamma_j^{1\ell}$
%
%
one gets the one-loop functional,
\begin{equation}
\label{eq.Wrenor}
W[J]\,\,=\,\,  \overline{S}_0^{\rm LO}[\Phi_k^{cl},J] \, \, + \,\, \overline{S}_1^{\rm LO}[\Phi_k^{cl},J]
\,\, \, +\, \, [...]\, .
\end{equation}
We have written here  $\overline{S}_0^{\rm LO}[\Phi_k^{cl},J]$ since
only the $\cO(p^2)$ operators in  $S_0^{\rm LO}[\Phi_k^{cl},J]$ are relevant for $\Pi(s)$ at tree-level.
Hence,  the replacement $W[J]\longrightarrow \overline{W}[J]$
leaves unchanged the part of the generating functional under study.
This operation is understood in the sense of removing any
operator in $W[J]$ of order higher than $p^2$ (contributing to $\Pi(s)$ or $\mM_\Pi$).
This removes the local terms $\mO_{c_j}[\pi^{cl},J]$
that produce the polynomial contributions in correlator.
The separation of the $\Gamma_j^{1\ell}$ is not arbitrary since these are the only
local operators in $S_1^{\rm LO}[\Phi_k^{cl},J]$ of order $p^4$ or higher.
This means that any local contribution $c_j$ or $\Gamma_j^{1\ell}$
appearing in the generating functional $W[J]$,
in the amplitudes $\Pi(s)$ or in related processes $\mM_\Pi$
can be directly truncated and removed from the computation at the very beginning.

After the transformation into $\overline{W}[J]$, one reaches
a functional that generates  the two-point Green-functions  and
is renormalizable; a NLO shift of the parameters
in the $\cO(p^2)$ LO action $\overline{S}_0^{\rm LO}[\Phi_k^{cl},J]$
is enough to absorb the UV divergences of
the one-loop contribution  $\overline{S}_1^{\rm LO}[\Phi_k^{cl},J]$.

Although \  \ this  \ argumentation \  \  only \
considers  \  \ spin--0 correlators and assumes unsubtracted dispersion relations,
it provides an interesting insight.
The local terms $\Gamma_j^{1\ell}$, generated
by the one-point Feynman integrals $A_0(M^2)$, may depend on the
realization of the hadronic lagrangian~\cite{rcht,Donoghue,HLS}.
However, this arbitrariness is not reflected in the physical amplitudes, which
are uniquely determined in terms of the two-meson absorptive contributions
Im$\Pi(t)_{_{\Phi_1\Phi_2}}$ and the resonance couplings
$F_k^r$ and  masses $M_k^r$.
Likewise, it is important to recall that on the contrary to what happens in momentum space,
the QCD correlators are finite in configuration space and no problem of definition  should arise
in the large--$N_C$ amplitudes.
For instance,  large--$N_C$ QCD holographic models do not find problems
on handling the infinite tower of states in the five-dimensional
configuration space but
the infinite summations of Kaluza-Klein modes in momentum space
may eventually become ill-defined~\cite{5D}.
Further investigations along these lines are
relegated to future works.

\section*{Conclusions}

The present note demonstrates
that the moments $\mA_\Pi^{(n)}(s)$
are  determined in QCD at NLO in $1/N_C$ by
the operators in $S_0^{\rm LO}[\Phi_k,J]$,
which give the two-meson absorptive contributions
Im$\Pi(t)_{_{\Phi_1\Phi_2}}$ and the renormalized couplings  $F_k^r$ and $M_k^r$.
These two quantities provide the residues $Z_k$ of the QCD matrix element and the
position of the poles at NLO in perturbation theory.
Moreover, it is shown that the full correlator  $\Pi(s)$ is actually determined
by the same parameters as $\mA_\Pi^{(n)}(s)$ together with one renormalization condition $\Pi(s_0)$
in the case of divergent QCD correlators.

In a general effective field theory, the NLO amplitude is determined by
new NLO terms in the action.
This analysis suggests that a large-$N_C$ resonance QFT should not be regarded as an effective theory.
In the case of two-point Green-functions,
it rather resembles  a ``renormalizable theory'' up operators
$c_j \mO_{c_j}[\pi,J]$
containing at most Goldstone fields and external sources.
The local couplings $c_j$ eventually cancel the one-loop terms $\Gamma_j^{1\ell}$
and become irrelevant for the amplitude,
which can be computed even if their value remains unknown.
Hence, they can be
dropped  from the calculation at the very beginning.

\section*{Acknowledgments}

I would like to thank S.~Peris, A.~Pich, I.~Rosell and
P.~Ruiz-Femen\'\i a  for their criticisms and comments on the manuscript.
The ideas in the present work were partially developed
at IPN-Orsay and during the visit to ECT* on July 2006.
  This work is supported in part by National
 Nature Science Foundations of China under contract number
 10575002,
  10421503


\end{document}